 \documentclass[journal]{IEEEtran}

\usepackage{graphicx}

\usepackage{graphics}
\usepackage{cite}
\usepackage{url}
\usepackage{enumerate}
\usepackage{latexsym}
\usepackage{newlfont}
\usepackage{amsthm}
\usepackage{amsmath}
\usepackage{amsfonts}
\usepackage{amssymb}
\usepackage{dropping}
\usepackage{pifont}
\usepackage[usenames,dvipsnames]{color}

\makeatletter

\begin{document}
\title{THRIVE: Threshold Homomorphic encRyption based secure and privacy preserving bIometric VErification system}

\newcommand{\mehmet}[1]{\authnote{Mehmet}{#1}}
\newcommand{\cagatay}[1]{\authnote{Cagatay}{#1}}
\def\msk#1{\textcolor{red}{[[#1]]}}
\def\ck#1{\textcolor{blue}{[[#1]]}}

\makeatletter \def\@r@al{\raise\arrowlower}
\def\@lrulefill{\cleaders\hbox{$\mkern-2mu\@r@al\hbox{$-$}\mkern-2mu$}\hfill}
\def\@lendofrule{\@r@al\hbox{$-$}\mkern-4mu}
\def\@rendofrule{\mkern-4mu\@r@al\hbox{$-$}}
\def\@rafill{\mkern-6mu\@lrulefill\@rendofrule\@r@al\llap{$\rightarrow$}}
\def\@lafill{\m@th\@r@al\rlap{$\leftarrow$}\@lendofrule\@lrulefill\mkern-6mu}
\def\@rdfill{\mkern-6mu\@lrulefill\@rendofrule}
\def\@ldfill{\m@th\@lendofrule\@lrulefill\mkern-6mu} \newdimen\minarrowwidth
\minarrowwidth=0pt \newdimen\arrowwidth \newdimen\arrowlower \arrowlower=-4pt
\def\@showmessage#1{\setbox\@tempboxa\hbox{$#1$}
\ifdim\wd\@tempboxa<\minarrowwidth \arrowwidth\minarrowwidth \else
\arrowwidth\wd\@tempboxa \fi \hbox{$\m@th\displaystyle\mathop{\hbox to
\arrowwidth {$\m@th\@lrulefill$}}\limits^{\hfill\box\@tempboxa\hfill}$}}
\def\sends{\@sends} \def\@sends#1{\@ldfill\@showmessage{#1}\@rafill}
\def\receives{\@receives} \def\@receives#1{\@lafill\@showmessage{#1}\@rdfill}
\makeatother

\newcommand{\ignore}[1]{}
\newcommand{\com}[3]{\textrm{commit}$_{#1}(#2;#3)$}
\newcommand{\comh}[4]{\textrm{commit}$_{#1}^{#2}(#3;#4)$}
\newcommand{\comnor}[2]{\textrm{commit}$_{#1}(#2)$}
\pagestyle{plain}

\author{\IEEEauthorblockN{
Cagatay~Karabat\IEEEauthorrefmark{1}'\IEEEauthorrefmark{2},
Mehmet Sabir Kiraz\IEEEauthorrefmark{1}, 
Hakan~Erdogan\IEEEauthorrefmark{2},
Erkay~Savas\IEEEauthorrefmark{2}\\
\IEEEauthorblockA{\IEEEauthorrefmark{1}TUBITAK BILGEM UEKAE, Kocaeli, Turkey} \\
\IEEEauthorblockA{\IEEEauthorrefmark{2}Sabanci University, TR-34956, Tuzla, Istanbul, Turkey}\\
\{cagatay.karabat, mehmet.kiraz\}@tubitak.gov.tr, \{haerdogan, erkays\}@sabanciuniv.edu}}


\maketitle

\begin{abstract}

In this paper, we propose a new biometric verification and template protection system which we call the THRIVE system. The system includes novel enrollment and authentication protocols based on threshold homomorphic cryptosystem where the private key is shared between a user and the verifier. In the THRIVE system, only encrypted binary biometric templates are stored in the database and verification is performed via homomorphically randomized templates, thus, original templates are never revealed during the authentication stage. The THRIVE system is designed for the malicious model where  the cheating party may arbitrarily deviate from the protocol specification. Since threshold homomorphic encryption scheme is used, a malicious database owner cannot perform decryption on encrypted templates of the users in the database. Therefore, security of the THRIVE system is enhanced using a two-factor authentication scheme involving the user's private key and the biometric data. We prove security and privacy preservation capability of the proposed system in the simulation-based model with no assumption. The proposed system is suitable for applications where the user does not want to reveal her biometrics to the verifier in plain form but she needs to proof her physical presence by using biometrics. The system can be used with any biometric modality and biometric feature extraction scheme whose output templates can be binarized. The overall connection time for the proposed THRIVE system is estimated to be 336~ms on average for 256-bit biohash vectors on a desktop PC running with quad-core 3.2 GHz CPUs at 10 Mbit/s up/down link connection speed. Consequently, the proposed system can be efficiently used in real life applications.

\end{abstract}


\IEEEpeerreviewmaketitle

\section{Introduction}

\IEEEPARstart{I}{n} recent years, public and commercial organizations invest on secure electronic authentication (e-authentication) systems to reliably verify identity of individuals. Biometrics is one of the rapidly emerging technologies for e-authentication systems \cite{Jain1}. However, it is impossible to discuss biometrics without security and privacy issues \cite{Prabhakar, Jain2}. Biometrics, which are stored in a smart card or a central database, is under security and privacy risks due to increased number of attacks against identity management systems in recent years \cite{Prabhakar, Ratha, Roberts, Jain2}.

Security and privacy concerns on biometrics limit their widespread usage in real life applications. The initial solution that occurs to mind for security and privacy problems is to use cryptographic primitives. On the other hand, biometric templates cannot be directly used with conventional encryption techniques (i.e. AES, 3DES) since biometric data are inherently noisy \cite{Kevenaar}. In other words, the user is not able to present exactly the same biometric data repeatedly. Namely, when a biometric template is encrypted during the enrollment stage, it should be decrypted to pass the authentication stage for comparison with the presented biometric. This, however, again leads to security and privacy issues for biometric templates at the authentication stage \cite{Kevenaar}.  Another problem with regards to such a solution is the key management, i.e. storage of encryption keys. When a malicious database manager obtains decryption keys, he can perform decryption and obtain biometric templates of all users. Similar problems are valid for cryptographic hashing methods. Since cryptographic hash is a one-way function, when a single bit is changed the hash sum becomes completely different due to the avalanche effect \cite{Feistel}. Thus, successful authentication by exact matching cannot be performed even for legitimate users due to the noisy nature of biometric templates. Therefore, biometric templates cannot directly be used with the cryptographic hashing methods.

Biometric systems which use error correction methods are proposed to cope with noisy nature of the biometric templates in the literature \cite{Hao, Kanade, Juels2}. In such systems, the biometric data collected at the enrollment stage is exactly the same with the biometric data collected at the authentication stage since they use error correction methods. In other words, these systems can get error-free biometric templates and thus cryptographic primitives (i.e. encryption and hashing) can successfully be employed without suffering from the avalanche effect \cite{Davida, Juels2, Tulyakov, Kevenaar}. However, high error correcting capability requirements make them impractical for real life applications \cite{Sutcu2}. Furthermore, side information (parity bits) is needed for error correction and this may lead to information leakage and even other attacks (i.e., error correcting code statistics, and non-randomness attacks) \cite{Stoianov1}. Zhou \textit{et al.} clearly demonstrate in their works that redundancy in an error correction code causes privacy leakage for biometric systems \cite{Zhou, Zhou2}.

Although biometric template protection methods are proposed to overcome security and privacy problems of biometrics \cite{Jain2, Cimato1, Matsumoto, Putte, Bringer, Barni, Nkumar, Feng, Bui, Juels, Karabat, Bai, Kuan, Rathgeb, Lumini}, recent research shows that security issues are still valid for these schemes \cite{Scheirer, Adler, Boult, Kong, Vielhauer, Cheung, Kummel}. Furthermore, there are a number of works on privacy leakages of biometric applications \cite {Simoens, Ignatenko1},  and yet more biometrics template protection methods \cite{Zhou, Zhou2, Simoens2}. In the literature, Zhou \textit{et al.} propose a framework for security and privacy assessment of biometric template protection methods \cite{Zhou}. In addition, Ignatenko \textit{et al.} analyze the privacy leakage in terms of the mutual information between the public helper data and biometric features in a biometric template protection method. A trade-off between maximum secret key rate and privacy leakage is given in their works \cite{Ignatenko1, Ignatenko2}. 

Recently, homomorphic encryption methods are used with biometric feature extraction methods to perform verification via encrypted biometric templates \cite{Erkin, Sadeghi, Barni, Barni2}. However, these methods offer solutions in the honest-but-curious model where each party is obliged to follow the protocol, but can arbitrarily analyze the knowledge that it learns during the execution of the protocol to obtain some additional information. The existing systems are not designed for the malicious model where each party can arbitrarily deviate from the protocol and may be corrupted. Moreover, they do not take into account security and privacy issues of biometric templates stored in the database \cite{Barni, Barni2}. The authors state that their security model will be improved in the future work by applying encryption methods also on the biometric templates stored in the database. Furthermore, some of these systems are just designed for a single biometric modality or a specific feature extraction method which also limits their application areas \cite{Erkin, Sadeghi}. In addition, an adversary can enroll himself on behalf of any user to their systems since they do not offer any solutions for malicious enrollment. Finally, all these systems suffer from computational complexity.

Biohashing schemes are one of the emerging biometric template protection methods \cite{Karabat, Bai, Kuan, Rathgeb, Lumini}. These schemes offer low error rates and fast verification at the authentication stage. However, they suffer from several attacks reported in the literature \cite{Kong, Vielhauer, Cheung, Kummel}. These schemes should be improved  to be safely used in a wide range of real life applications. In our work, we develop new enrollment and authentication protocols for biometric verification methods. Our goal is to increase security and enhance privacy of the biometric schemes. The THRIVE system can work with any biometric feature extraction scheme whose outputs are binary or can be binarized. Since biohashing schemes can output binary templates called a biohash, they can be successfully used with the proposed system.

\subsection{Our contributions}

In this paper, we address adversary attacks in case of an active attacker who aims to gain access to the system in the malicious attack model. By taking these adversary attacks into account, we develop a new biometric authentication system based on threshold homomorphic cryptosystem. Our main goal is to increase the security of the system and preserve privacy of biometric templates of the users. The contributions of this work can be summarized as follows:

\begin{itemize}
\item A new biometric authentication system (which we call the THRIVE system) is proposed in the malicious model and the proposed system can be used with any existing biometric modality whose output can be binarized. 

\item Even if an adversary gains an access to the database and steals encrypted biometric templates,  he can neither authenticate himself by using these encrypted biometric templates due to the proposed authentication protocol, nor decrypt these encrypted biometric templates due to the $(2,2)$-threshold homomorphic encryption scheme. 

\item Only encrypted binary templates are stored in the database and biometric templates are never released even during authentication. Thus, the proposed system offers a new and advanced biometric template protection method without any helper data. In addition, only legitimate users can enroll in the proposed system since a signature scheme is used with the proposed enrollment protocol.

\item The THRIVE system can be used in the applications where the user does not trust the verifier since she does not need to reveal her biometric template and/or private key to authenticate herself and  the verifier does not need to reveal any data to the user with the proposed authentication protocol. 

\item Even if an adversary intercepts the communication channel between the user and the verifier, he cannot obtain any useful information on the biometric template since all exchanged messages are randomized and/or encrypted and he cannot perform decryption due to the $(2,2)$-threshold homomorphic encryption scheme. Furthermore, he cannot use the obtained data from message exchanges in this communication channel since nonce and signature schemes are used together in the authentication.

\item The THRIVE system is a two-factor authentication system (biometric and secret key) and is secure against illegal authentication attempts. In other words, a malicious adversary cannot gain access to the proposed system without having the biometric data and the private key of a legitimate user by performing adversary attacks described in \cite{Ratha2} as well as hill-climbing attacks \cite{Adler2, Galbally, Uhl, Martinez}.

\item In the THRIVE system, the generated protected biometric templates are irreversible since templates are encrypted they are irreversible by definition as soon as decryption key is not stolen. 

\item The THRIVE system can generate a number of protected templates from the same biometric data of a user due to the randomized encryption and biohashing. Thus, it ensures diversity. Besides, they are also cancelable i.e. when they are stolen, they can be re-generated.

\item A THRIVE  authentication protocol run requires only 336 ms on average for 256 bit biohash vectors and 671 ms on average for 512 bit biohash vectors on a desktop PC with quad-core 3.2 GHz CPUs at 10 Mbit/s up/down link connection speed. Therefore, the proposed system is sufficiently efficient to be used in real-world applications.

\end{itemize}

The paper is structured as follows. Related work is addressed in Section 2. Preliminaries are described in Section 3. The proposed biometric authentication system is introduced in Section 4. Security proof of the proposed protocols are given in Section 5. Complexity analysis of the proposed system is discussed in Section 6.  Section 7 concludes the paper.

\section{Related Work}

Biometric template protection schemes are proposed to mitigate the security and privacy problems of biometrics \cite{Jain2, Matsumoto, Putte, Bringer, Barni, Nkumar, Feng, Bui, Juels}. However, various vulnerabilities of these technologies are reported in the literature \cite{Scheirer, Adler, Boult}. Jain \textit{et al.} classify biometric template protection schemes into two main categories \cite{Jain2}: 1) Feature transformation based schemes, 2) Biometric cryptosystems as illustrated in Figure~\ref{categorize}. 

The main idea behind biometric cryptosystems (also known as biometric encryption systems) is either binding a cryptographic key with a biometric template or generating the cryptographic key directly from the biometric template~\cite{Uludag}. Thus, the biometric cryptosystems can be classified into two main categories: 1) Key binding schemes, 2) Key generation schemes. The biometric cryptosystems use helper data, which is public information, about the biometric template for verification. Although helper data are supposed not to leak any critical information about the biometric template, Rathgeb \textit{et al.} show that helper data is vulnerable to statistical attacks \cite{Rathgeb2}. Furthermore, Ignatenko \textit{et al.} show how to compute a bound on possible secret rate and privacy leakage rate for helper data schemes ~\cite{Ignatenko3}. Adler performs hill-climbing attack against biometric encryption systems ~\cite{Adler}. In addition, Stoianov \textit{et al.} propose several attacks (i.e., nearest impostors, error correcting code statistics, and non-randomness attacks) to biometric encryption systems \cite{Stoianov1}. 

In the literature, fuzzy commitment \cite{Juels2} and fuzzy vault schemes \cite{Juels} are categorized under the key binding schemes. These schemes aim to bind a cryptographic key with a biometric template. In ideal conditions, it is infeasible to recover either the biometric template or the random bit string without any knowledge of the user's biometric data. However, this is not the case in reality because biometric templates are not uniformly random. Furthermore, error correction codes (ECC) used in biometric cryptosystems lead to statistical attacks (i.e., running ECC in a soft decoding or erasure mode and ECC Histogram attack) \cite{Stoianov1, Stoianov2}. Ignatenko \textit{et al.} show that fuzzy commitment schemes leak information in cryptographic keys and biometric templates which lead to security flaws and privacy concerns \cite{Ignatenko1, Ignatenko2}. In addition, Zhou \textit{et al.} argue that fuzzy commitment schemes leak private data. Chang \textit{et al.} describe a non-randomness attack against fuzzy vault scheme which causes distinction between the minutiae points and the chaff points \cite{Chang}. Moreover, Kholmatov \textit{et al.} perform a correlation attack against fuzzy vault schemes \cite{berrin}.

Keys are generated from helper data and a given biometric template in key generation schemes \cite{Jain2}. Fuzzy key extraction schemes are classified under key generation schemes and use helper data \cite{Dodis1, Dodis2, Yagiz, Ong, Arakala}. These schemes can be used as an authentication mechanism where a user is verified via her own biometric template as a key. Although fuzzy key extraction schemes provide key generation from biometric templates, repeatability of the generated key (in other words stability) and the randomness of the generated keys (in other words entropy) are two major problems of them \cite{Jain2}. Boyen  \textit{et al.} describe several vulnerabilities (i.e. improper fuzzy sketch constructions may lead information on the secret, biased codes may cause majority vote attack, and permutation leaks) of the fuzzy key extraction schemes from outsider and insider attacker perspectives \cite{Boyen}. Moreover, Li \textit{et al.} mention that when an adversary obtains sketches, they may reveal the identity of the users \cite{Li}.

Biohashing schemes are simple yet powerful biometric template protection methods \cite{Karabat, Bai, Kuan, Rathgeb, Lumini} which can be classified under salting based schemes. It is worth pointing out that biohashing is completely different from cryptographic hashing. Although biohashing schemes are proposed to solve security and privacy issues, there are still security and privacy issues associated with them \cite{Kong, Vielhauer, Cheung, Kummel}. In these works, the authors claim that biohashes can be reversible under certain conditions and an adversary can estimate biometric template of a user from her biohash. Consequently, when biohashes are stored in the databases and/or smart cards in their plain form, they can threaten the security of the system as well as the privacy of the users. Moreover, an adversary can use an obtained biohash to threaten the system security by performing malicious authentication. Furthermore, when the secret key is compromised, an adversary can recover the biometric template since these schemes are generally invertible \cite{Jain2}. 

Non-invertible transform based schemes use a non-invertible transformation function, which is a one-way function, to make the biometric template secure \cite{Sutcu, Jin, Yang}. User's secret key determines the parameters of  non-invertible transformation function and this secret key should be provided at the authentication stage. Even if an adversary obtains the secret key and/or the transformed biometric template, it is computationally hard to recover the original biometric template. On the other hand, these schemes suffer from the trade-off between discriminability and non-invertibility which limits their recognition performance \cite{Jain2}.

Another security approach is the use of cryptographic primitives (i.e. encryption, hashing) to protect biometric templates. These works generally focus on fingerprint-based biometric systems. Tuyls \textit{et al.} propose the fingerprint authentication system which incorporates cryptographic hashes \cite{Tuyls}. They use an error correction scheme to get exactly the same biometric template from the same user in each session which is similar to the fuzzy key extraction schemes. They store cryptographic hashes of biometric templates in the database and make comparison in the hash domain. However, there is no guarantee to get exactly the same biometric templates from the user even if the system incorporates an error correction scheme in real life applications since it is limited with the pre-defined threshold of error correction capacity. They also use helper data which are sent over a public channel and this may lead to security flaws as well. Moreover, an adversary can threaten the security of the system when he performs an attack against the database since he can obtain the user id, helper data and the hashed version of the secret which is generated by the biometric data and the helper data. Although the adversary cannot obtain the biometric data itself in its plain form, he can get all needed credentials (i.e. hash values of the secrets) to gain access to the system.

Kerschbaum \textit {et al.} propose a protocol to compare fingerprint templates without actually exchanging them by using secure multi-party computation in the honest-but-curious model \cite{Kerschbaum}. At the enrollment stage, the user gives her fingerprint template, minutiae pairs and PIN to the system. Thus, the verifier knows the fingerprint templates which are collected at the enrollment stage.  Although the user does not send her biometric data at the authentication, the verifier already has the user's enrolled biometric data and this threatens the privacy of the user in case of a malicious verifier. In addition, a malicious verifier can use these fingerprint templates for malicious authentication. Furthermore, since the fingerprint comparison reveals the  matching scores (i.e. Hamming distance\cite{Hamming}), the attacker can perform a hill climbing attack against this system. Apart from these security and privacy flaws, the authors just focus on secure comparison in their protocol and they do not develop any solutions for the malicious model. 

Erkin \textit {et al.} \cite{Erkin} propose a privacy preserving face recognition system for the eigen-face recognition algorithm \cite{Turk}. They design a protocol that performs operations on encrypted images by using the Pailler homomorphic encryption scheme. Later, Sadeghi \textit{et al.}  improve the efficiency of this system \cite{Sadeghi}. In both works, they use the eigen-face recognition algorithm together with homomorphic encryption schemes. However, they limit the recognition performance of the system with the eigen-face method although there are various feature extraction methods which perform better than it. Unfortunately, their system cannot be used for any other feature extraction method for face images. Moreover, they do not use a threshold cryptosystem which prevents from a malicious party aiming to perform decryption by himself. Storing face images (or corresponding feature vectors) in the database in plain is the most serious security flaw of this system. An adversary, who gains access to the database, can obtain all face images. Therefore, the adversary can perform the sixth attack type - attack against the database which definitely threatens the security of the system and the privacy of the users.

Barni \textit {et al.} \cite{Barni, Barni2} propose a privacy preservation system for fingercode templates by using homomorphic encryption in the honest-but-curious model. They, however, do not propose any security and privacy solutions on the biometric templates stored in the database. This issue is mentioned as a future work in their paper. In addition, they do not use threshold encryption which would prevent from  a malicious party aiming to perform decryption by himself. Therefore, their proposed system is open to adversary attacks against the database as stated in their work. They do not address the malicious enrollment issue as well. Moreover, the user must trust the server in their system. Although they achieve better performance than \cite{Erkin, Sadeghi} in terms of bandwidth saving and time efficiency, they do not address the applications where the user and the verifier do not trust each other (e.g. the malicious model).

There are also some works on secure Hamming distance calculation by using cryptographic primitives \cite{Osadchy, Rane, BringerCP13, Kulkarni}. These papers, however, limit their works only with secure Hamming distance calculation. These methods do not address biometric authentication as a whole and fails to satify security, privacy, template protection at the same time by taking into account computational efficiency which is very critical for real-world applications. Osadchy \textit {et al.} \cite{Osadchy} propose Pailler homomorphic encryption based secure Hamming distance calculation for face biometrics. The system is called SCiFI. Although they claim that SCiFI is computationally efficient, it mostly uses pre-computation techniques. Its pre-computation time includes processing time that must be done locally by each user before using the system each time. They report that SCiFI's online running time takes 0.31 seconds for a face vector of size 900 bits however its offline computation time takes 213 seconds. Since these computations should be done by the user just before each attempt to use the system, the protocol is not that much efficient. Besides, SCiFI is only secure for semi-honest adversaries. Rane \textit {et al.} \cite{Rane} also propose secure Hamming distance calculation for biometric applications. However, their proposed method fails to ensure biometric database security since biometric templates are stored in plain format in the database. Thus, a malicious verifier can threaten a user's security and privacy. Bringer \textit {et al.} \cite{BringerCP13} propose a secure Hamming distance calculation for biometric application. The system is called SHADE and it is based on committed oblivious transfer \cite{KSV06}. However, they also cannot guarantee biometric database security since biometric templates are stored in plain form in the database. Kulkarni \textit {et al.} \cite{Kulkarni} propose a biometric authentication system based on \textit{somewhat} homomorphic encryption scheme of  Boneh \textit{et al.}\cite{BGN06} which allows an arbitrary number of addition of ciphertexts but supports only one multiplication operation between the ciphertexts. Although the values stored on the enrollment server are the XORed values of the biometric template vector with the corresponding user's key, the user first extracts and sends her biometric features to the trusted enrollment server. Again this system uses a trusted enrollment server and fails to protect security and to preserve privacy of a user against a malicious database manager. In addition, the system is not efficient since 58 sec are required for successful authentication of a 2048 bit binary feature vector.

\section{Preliminaries}

\subsection{Threshold Homomorphic Cryptosystem}

In this section, we briefly describe underlying cryptographic primitives of the protocols. Given a public key encryption scheme, let $m \in \mathcal{M}$ denote its message or plaintext space, $c \in \mathcal{C}$ the ciphertext space, and $r \in \mathcal{R}$ its randomness. Let $c=\text{\sf Enc}_{pk}(m;r)$ depict an encryption of $m$ under the public key $pk$ where $r$ is a random value. Let $sk$ be its corresponding private key, which allows the holder to retrieve a message from a ciphertext. The decryption is done with the private key $sk$ as $m=\text{\sf Dec}_{sk}(c)$.

In a $(t,n)$-threshold cryptosystem, the knowledge of a private key is distributed among parties $P_1,\ldots, P_n$. Then, at least $t$ of these parties are required for successful decryption. On the other hand, there is a public key to perform encryption. More formally, let $P_1,\ldots,P_n$ be the participants. We define a $(t,n)$-threshold encryption scheme with three phases as follows:

\begin{itemize}
  \item In the \emph{\textbf{key generation}} phase, each participant $P_i$ receives a pair $(pk_i,sk_i)$, where $pk_i$ and $sk_i$ are the \emph{shares} of the public and secret key, respectively. Then, the overall public key $pk$ is constructed by collaboratively \emph{combining} the shares. Finally $pk$ is broadcast to allow anyone to encrypt messages in $\mathcal{M}$. The shares of this public key are also broadcast which allow all parties to check the correctness of the decryption process.
  
  \item The \emph{\textbf{encryption}} phase is done as in any public key encryption cryptosystem. If $m\in\mathcal{M}$ is the message, a (secret) random value $r$ from $\mathcal{R}$ is chosen and $c=\text{\sf Enc}_{pk}(m;r)$ is broadcast under a public key $pk$.
	
  \item In the \emph{\textbf{threshold decryption}} phase, given that $t$ (or more) participants agree to decrypt a ciphertext $c$, they follow two steps. First, each participant produces a decryption share by performing $S_{i}^j=\text{\sf Dec}_{sk_{i}^j}(c)$, $j=1,\ldots,t$. After broadcasting $S_i^j$, they all can apply a reconstruction function $\mathcal{F}$ on these shares so that they can recover the original message by performing $m=\mathcal{F}(S_{i}^1,\ldots,S_{i}^t)$ where $P_{i}^1,\ldots,P_{i}^t$ represent the group of $t$ participants willing to recover $m$. 
\end{itemize}

In case of a $(t,n)$-threshold scheme, the additional requirement is that if less than $t$ parties gather their correct shares of the decryption of a given ciphertext, they will get no information whatsoever about the plaintext. In the proposed system, we use the $(2,2)$-threshold cryptosystem between the claimer (the user) and the verifier where both players must cooperate to decrypt. 

A public key encryption scheme is said to be additively homomorphic if given $c_1=\text{\sf Enc}(m_1;r_1)$ and $c_2=\text{\sf Enc}(m_2;r_2)$ it follows that $c_1c_2=\text{\sf Enc}(m_1 + m_2;r_3)$ where $m_1, m_2 \in \mathcal{M}$ and $r_1, r_2, r_3 \in \mathcal{R}$. There are various versions of threshold homomorphic cryptosystems. The most widely used are ElGamal \cite{Elgamal} or Paillier \cite{Paillier} cryptosystems. In our proposal, we will use a threshold version of Goldwasser-Micali (GM) encryption scheme (i.e., between a user and a verifier) proposed by Katz and Yung in \cite{KY02}. Note that GM scheme is XOR-homomorphic \cite{GM84}, i.e., given any two bits $b_1, b_2$ in $\{0,1\}$, any random values $r_1$, $r_2$ $\in$ $\mathcal{R}$ , and any encryptions $\text{\sf Enc}(b_1,r_1), \text{\sf Enc}(b_2,r_2)$, it is easy to compute $\text{\sf Enc}(b_1 \oplus b_2, r_1 r_2)$. 

In the proposed protocol, we use a variant of the threshold decryption protocol which is the so-called private threshold decryption \cite{Damgard}. The requirement of this protocol is that one of the $t$ parties will be the only party who will recover the secret. All $t-1$ other parties follow the protocol and broadcast their shares to achieve this requirement. The party who will learn the plaintext proceeds with the decryption process privately, collects all decryption shares from the $t-1$ other parties, and privately reconstructs the message. The remaining parties will not get any information about this message.

\subsubsection{Threshold XOR-Homomorphic Goldwasser-Micali Encryption Scheme}
We next give a brief explanation of (2,2)-GM cryptosystem between two users (in our proposal, between a user and a verifier) using a Trusted Dealer. We note that one can also exclude a trusted dealer using the scheme in \cite{KY02}:\\
\textbf{Key generation:}\\
The trusted dealer first chooses prime numbers $p$ and $q$ $(\|p\| = \|q\| = n)$ such that $N=pq$ and $p$ $\equiv$ $q$ $\equiv$ $3$ $\mod 4$. The dealer next chooses $p_1$, $q_1$, $p_2$, $q_2$ $\in_R$ $(0,2^{2n})$ such that $p_1 \equiv q_1 \equiv 0 \mod 4$ and $p_2 \equiv q_2 \equiv 0 \mod 4$. He sets $p_0$ = $p - p_1 - p_2$ and $q_0$ $=$ $q - q_1 -q_2$ and sends $(p_1,q_1)$ to the first party and $(p_2,q_2)$ to the second party. He finally broadcasts $(p_0, q_0, N)$.\\
\textbf{Encryption of a bit $b$ $\in$ $\{0,1\}$:}\\
Choose $r$ $\in_R$ $\mathbb{Z}_N$ and compute a ciphertext $C = (-1)^b r^2 \mod N$.\\
\textbf{Decryption:}\\
All parties compute the Jacobi symbol $J= (\frac{C}{N})$. If $J \neq 1$ then all parties stop because either the encryption algorithm was not run honestly or the ciphertext was corrupted during the transmission. (Note that $(\frac{C}{N})$ is always 1, because $(\frac{C}{N})$ = $(\frac{C}{p})$ $(\frac{C}{q})$= 1 (i.e., either $(\frac{C}{p})=1$ and $(\frac{C}{q})=1$ or $(\frac{C}{p})=-1$ and $(\frac{C}{q})=-1)$). If $J=1$ then the first party broadcasts $b_1 = C^{{(-p_1-q_1)}/4} \mod N$. The second party (who is going to decrypt) will privately compute $b_0= C^{{(N -p_0 -q_0 +1)}/4} \mod N$ and $b_2 = C^{{(-p_2-q_2)}/4} \mod N$. Finally, the decrypted bit $b$ is computed as $b = (1-b_0 b_1 b_2 \mod N)/2$. 

Note that it is easy to see whether $C$ is a quadratic residue by computing $b\equiv C^{(N-p-q-+1)/4}$ $\mod$ $N$. The reason is briefly as follows. We first note that by Euler's theorem $C^{\phi(N)} \equiv 1 \mod N$ where $\phi(N)$ $=$ $(p-1)(q-1)$. We also know that $C$ is quadratic residue iff $C^{\phi(N)/2} \equiv 1 \mod N$. If the Jacobi symbol $J$ $=$ $(\frac{C}{N})$ = 1 then by using $(\frac{C}{p})$ $(\frac{C}{q})$ $=$ $1$ we have either $(\frac{C}{p})=1$ and $(\frac{C}{q})=1$ or $(\frac{C}{p})=1$ and $(\frac{C}{q})=1$. If $(\frac{C}{p})$ $=$ $1$ (resp. -1) and $(\frac{C}{q})$ $=$ $1$ (resp. -1) then $C^{{p-1}/2}$ $\equiv$ 1 $\mod p$ (resp. $-1 \mod p$) and $C^{{q-1}/2}$ $\equiv$ $1$ $\mod q$ (resp. $-1 \mod q$). Hence, for both cases $C^{(p-1)(q-1)/4}$ $\equiv$ 1 $\mod p$ and $C^{(p-1)(q-1)/4}$ $\equiv$ $1$ $\mod q$. By the Chinese Remainder Theorem, we have $C^{{(p-1)(q-1)}/4}$ $\equiv$ $1$ $\mod N$. Hence, $C$ is quadratic residue iff $b=1$.

\subsection{Biometric Verification Scheme}

Biometric verification schemes perform an automatic verification of a user based on her specific biometric data (e.g., face, fingerprint, iris). They have two main stages: 1) Enrollment stage, and 2) Authentication stage. The user is enrolled to the system at the enrollment stage. Then, she again provides her biometric data to the system at the authentication stage to prove her identity. Any biometric scheme, which provides binary outputs or whose outputs can be binarized, can work with the proposed threshold homomorphic cryptosystem. The THRIVE system can work with any biometric feature extraction method which produces fixed size vectors as templates and perform verification with distance calculations between the enrolled and the provided template at the authentication stage. When the output of a biometric feature extraction method is not binary, locality sensitive hashing can be used to binarize the feature vector \cite{gionis}. After binarization, the binary templates can successfully be used with the proposed system. In this paper, we use biohashing as an example algorithm for extracting binary biometric templates. Although biohashing has its own security and privacy preservation mechanism, we do not rely on these for the security or the privacy preservation features. Thus it can be replaced with any other binary feature extraction method.

Biohashing schemes are simple yet powerful biometric template protection methods \cite{Karabat, Bai, Kuan, Rathgeb, Lumini}. Biohash is a binary and pseudo-random representation of a biometric template. Biohashing schemes use two inputs: 1) Biometric template, 2) User's secret key. A biometric feature vector is transformed into a lower dimension sub-space using a pseudo-random set of orthogonal vectors which are generated from the user's secret key. Then, the result is binarized to produce a pseudo-random bit-string which is called the biohash. In an ideal case, the Hamming distance between the biohashes belonging to the biometric templates of the same user is expected to be relatively small. On the other hand, the distance between the biohashes belonging to different users is expected to be sufficiently high to achieve higher recognition rates.  

We descibe the random projection (RP) based biohashing scheme proposed by Ngo \textit{et al.} \cite{Ngo}. In this scheme, there are three main steps: 1) Feature extraction, 2) Random projection, 3) Quantization. These steps are explained for face biometrics.

\subsubsection{Feature Extraction}

The feature extraction is performed on the face images, which are collected at the enrollment stage, belonging to the users, $\textbf{I}_{i,j}\in\Re^{m\times n}$ where $i=1,\ldots,n$ and $n$ denotes number of users, $j=1,\ldots,L$ and $L$ denotes number of training images per user. The face images are lexicographically re-ordered and the training face vectors, $\textit{\textbf{x}}_{i,j}\in\Re^{(mn)\times1}$, are obtained. Then, Principle Component Analysis (PCA) \cite{Turk} is applied. 

\begin{equation}
 \textsl{\textbf{y}}_{i,j}=\textbf{A}(\textit{\textbf{x}}_{i,j}-\textit{\textbf{w}})
\end{equation}
where $\textbf{A}\in\Re^{k\times(mn)}$ is the PCA matrix trained by the face images in the training set, \textit{\textbf{w}} is the mean face vector, and $\textbf{\textsl{y}}_{i,j}\in\Re^{k\times1}$ is vector containing PCA coefficients belonging to the $j^{th}$ training image of the $i^{th}$ user.

\subsubsection{Random Projection}

At this phase, a RP matrix, $\textbf{R}\in\Re^{\ell\times k}$, is generated to reduce the dimension of the PCA coefficient vectors. The RP matrix elements are independent and identically distributed (\textit{i.i.d}) and generated from a Gauss distribution with zero mean and unit variance by using a Random Number Generator (RNG) with a seed derived from the user's secret key. The Gram-Schmidt (GS) procedure is applied to obtain an orthonormal projection matrix $\textbf{R}_{GS}\in\Re^{\ell \times k}$ to have more distinct projections. Finally, PCA coefficients are projected onto a lower $\ell$-dimensional subspace.

\begin{equation}
 \textsl{\textbf{z}}_{i,j}=\textbf{R}_{GS} \textit{\textbf{y}}_{i,j}
\end{equation}
where $\textbf{\textsl{z}}_{i,j}\in\Re^{\ell\times1}$ is an intermediate biohash vector belonging to the $j^{th}$ training image of the $i^{th}$ user.

\subsubsection{Quantization}

At this phase, the intermediate biohash vector $\textit{\textbf{z}}_{i,j}$ elements are binarized with respect to the threshold.

\begin{equation}
\lambda_{i,j}^k= 
\left\{\begin{array}{rcl}
1 &\mbox{if} & \textit{z}_{i,j}^k \geq \beta \\ 
0 & $Otherwise.$ \\
\end{array} \right.
\end{equation} 
where $\lambda_{i,j}\in\ \left\{0,1\right\}^{\ell}$ denotes biohash vector of the $j^{th}$ training image of the $i^{th}$ user and $\beta$ denotes the mean value of the intermediate biohash vector $\textit{z}_{i,j}$. 

A biohash vector, $\textbf{\textit{B}}_{enroll_i}$, for the $i^{th}$ user is stored in the database at the enrollment stage for verification purpose during the authentication stage. Note that, $\textbf{\textit{B}}_{enroll_i}$ can be any vector among $\lambda_{i,j}$ vectors in a real-world application. For simulation purposes, we take into account all possible biohashes for a user by computing $\lambda_{i,j}$. The user is authenticated if the Hamming distance between $\textbf{\textit{B}}_{enroll_i}$ and $\textbf{\textit{B}}_{auth_i}$  is below a threshold $\mu$.

\begin{equation}
\sum^{n}_{k=1}\textit{B}_{enroll_i}^k\oplus\textit{B}_{auth_i}^k\leq\mu
\end{equation}
where $\textit{B}_{enroll_i}^k$ denotes the $k^{th}$ bit of $\textbf{\textit{B}}_{enroll_i}$, $\textit{B}_{auth_i}^k$ denotes the $k^{th}$ bit of $\textbf{\textit{B}}_{auth_i}$, and $\oplus$ denotes the binary XOR (exclusive OR) operator. Consequently, the verifier decides whether the claimer is a legitimate user or not according the threshold.

\begin{figure*}[tb]
\centering
\begin{center}
\includegraphics [scale=0.43]{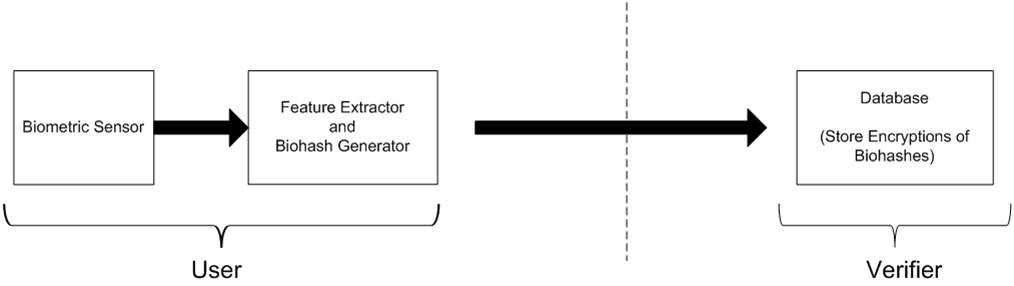}
\end{center}
\caption{Illustration of the THRIVE enrollment stage: the user has control over the biometric sensor, the feature extractor and the biohash generator whereas the verifier has control over the database.}
\label{enrollment_fig}
\end{figure*}

\begin{figure*}[htp]
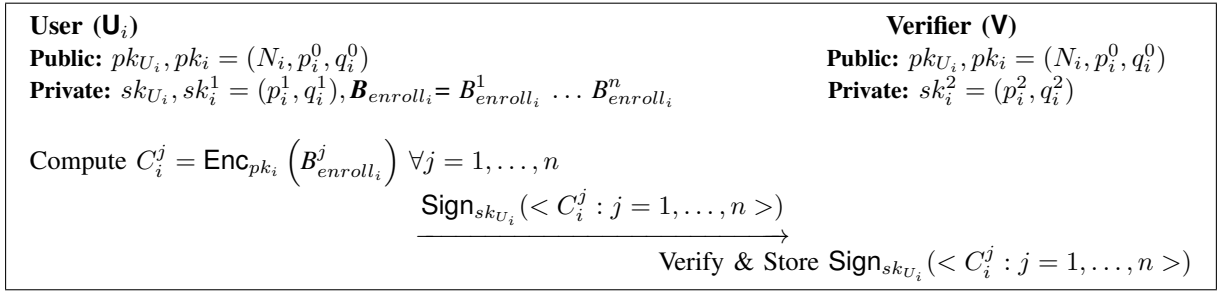

\centering
\fbox{\begin{tabular}{lcl}
\textbf{User ({\sffamily U}$_i$)} & \hspace{-1.6 cm} \textbf{Verifier ({\sffamily V})}\\
\textbf{\small{Public:}} $pk_{U_i}, pk_i= (N_i, p_i^0,q_i^0)$ \hspace{4 cm}  &\hspace{-0.4 cm} \textbf{\small{Public:}} $pk_{U_i}, pk_i= (N_i, p_i^0,q_i^0)$ \\
\textbf{\small{Private:}} $sk_{U_i}, sk_i^1=(p_i^1,q_i^1), \textbf{\textit{B}}_{enroll_i}$= ${\textit{B}}^1_{enroll_i}$ $\ldots$ ${\textit{B}}^n_{enroll_i}$ & \hspace{-1.6 cm} \textbf{\small{Private:}} $sk_i^2=(p^2_i,q^2_i)$\\\\
Compute $C^{j}_{i}=\text{\sf Enc}_{pk_i}\left({\textit{B}}^j_{enroll_i} \right)$  $\forall j=1, \ldots, n$\\

\hspace{5cm} $\sends{$\text{\sf Sign}$_{sk_{U_i}}(<C^{j}_{i}: j =1, \ldots, n>)}$\\
& \hspace{-2.3cm} Verify \& Store $$\text{\sf Sign}$_{sk_{U_i}}(<C^{j}_{i}:j =1,\ldots, n>)$ 
\end{tabular}}
\caption{The Proposed Enrollment Protocol}
\label{enrollment_protocol}
\end{figure*}

\section{The Proposed Biometric Authentication System}

In this section, the proposed biometric authentication system is introduced. In the proposed system, there are two major roles: 1. \emph{User} ({\sffamily $U_i$}) and 2. \emph{Verifier} ({\sffamily V}). The user has control of the biometric sensor, the feature extractor, and the biohash generator whereas the verifier has control of the database and the matcher. We assume that there is a trusted third party (TTP) which initially sets up the system public/private keys.

The TTP distributes the keys in the proposed system. There are public-private key pairs $(pk_i, (sk_i^1, sk_i^2))$ which are shared between the user and the verifier. $pk_i$ is the public key of the $i^{th}$ user, $U_i$, and both the user and the verifier have it. Recall that, when an enrollment biometric template is encrypted by $pk_i$, this can solely be decrypted using the private key shares of the user $(sk_i^1)$ and the verifier $(sk_i^2)$ collaboratively since the proposed system  is based on the $(2,2)$-threshold homomorphic cryptosystem. Here, $sk_i^1$ is the private key share of the $i^{th}$ user, $U_i$, and $sk_i^2$ is the private key share of the verifier.  Besides, there is a public-private key pair $(pk_{U_i}, sk_{U_i})$ which belongs to the $i^{th}$ user, $U_i$, where $pk_{U_i}$ is the public key and $sk_{U_i}$ is its associated private key to perform the signature operation. The verifier also has the public key $pk_{U_i}$ of the $i^{th}$ user, $U_i$.

\subsection{Enrollment Stage}

At this stage, the $i^{th}$ user $U_i$ has control over the biometric sensor, the feature extractor, and the biohash generator whereas the verifier has control over the database as  illustrated in Figure~\ref{enrollment_fig}. It is worth mentioning that biometric sensor authentication must be achieved in the proposed system before executing the enrollment protocol to prevent unauthorized sensors to be used as clients in the system by malicious users. This, however, is not explicitly indicated in the protocol in order not to clutter the paper. The proposed enrollment protocol is illustrated in Figure~\ref{enrollment_protocol} and steps of it are introduced as follows:

\begin{enumerate}

\item \textbf{Step 1}: The $i^{th}$ user, $U_i$, computes her biohash, $\textbf{\textit{B}}_{enroll_i}$ = ${\textit{B}}^1_{enroll_i}$ $\ldots$ ${\textit{B}}^n_{enroll_i}$ where ${\textit{B}}^j_{enroll_i}$ $\in$ $\{0,1\}$, $j=1,\ldots,n$. Next, the user encrypts her biohash, $C^{j}_{i} = $\text{\sf Enc}$_{pk_i}\left({\textit{B}}^j_{enroll_i}\right)$ $\forall j=1,\ldots,n$, by using the public key $pk_i$. Then, the user signs her encrypted biohash, \text{\sf Sign}$_{sk_{U_i}}(<C^{j}_{i}:j=1,\ldots,n>)$, and sends it to the verifier. 

\item \textbf{Step 2}: The verifier $V$ verifies \text{\sf Sign}$_{sk_{U_i}}(<C^{j}_{i}:j=1,\ldots,n>)$ by using $pk_{U_i}$ and stores the signature and encrypted biohash in the database. These data will be used for verification at the authentication stage. 
\end{enumerate}

Note that the proposed enrollment protocol uses the $(2,2)$-threshold homomorphic cryptosystem. Namely, both the user and the verifier have to cooperate to decrypt a ciphertext. Furthermore, the signature ensures that the data stored in the database are generated by a legitimate user.  

\textbf{Lemma 1.} \textit{Biohashes are not revealed at the enrollment stage.}

\textbf{\textit{Proof.}} (Sketch) At the enrollment stage, the $i^{th}$ user $U_i$ first encrypts her biohash and then signs it. After these computations, $U_i$ sends her encrypted and signed biohash \text{\sf Sign}$_{sk_{U_i}}(<C^{j}_{i}:j=1,\ldots,n>)$ to the verifier. Since the user's biohash is not sent in plain form, biohashes are not revealed to the verifier at the enrollment stage.

\textbf{Lemma 2.} \textit{An adversary cannot register as a legitimate user at the enrollment stage.}

\textbf{\textit{Proof.}} (Sketch) At the enrollment stage, the $i^{th}$ user $U_i$ encrypts her biohash by using the public key $pk_i$ and then signs her encrypted biohash by using her private key $sk_{U_i}$. Thus, $U_i$ sends encrypted and signed biohash \text{\sf Sign}$_{sk_{U_i}}(<C^{j}_{i}:j=1,\ldots,n>)$ to the verifier. The verifier knows $pk_{U_i}$ of the users. Since the verifier verifies the signature of the user, an adversary cannot register himself as a genuine user without having the private key of her $sk_{U_i}$ for computing \text{\sf Sign}$_{sk_{U_i}}(<C^{j}_{i}:j=1,\ldots,n>)$.


\subsection{Authentication Stage}

\begin{figure*} [tb]
\begin{center}
\includegraphics [scale=0.43]{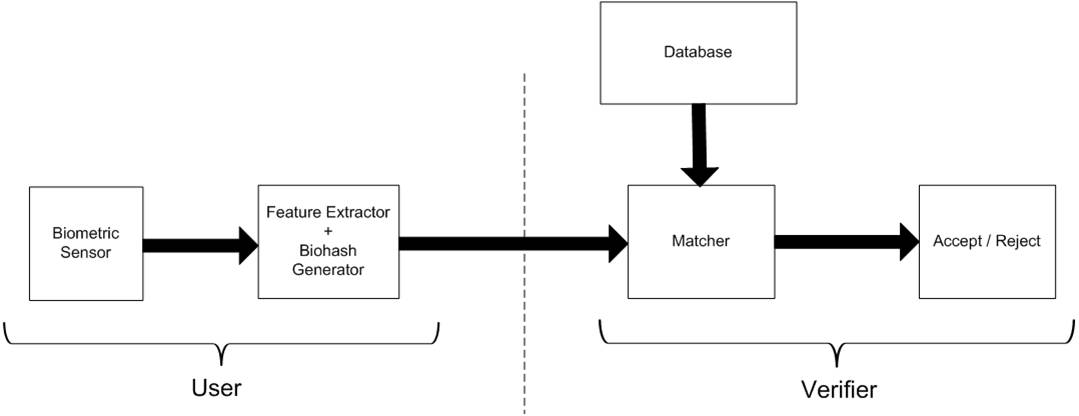}
\end{center}
\caption{Illustration of the THRIVE authentication stage: the user has control over the biometric sensor, the feature extractor and the biohash generator whereas the verifier has control over the database, the matcher and the decision maker.}
\label{authentication_fig}
\end{figure*}

At this stage, the $i^{th}$ user $U_i$, has control over the biometric sensor, the feature extractor, and the biohash generator whereas the verifier has control over the database, the matcher and the decision maker as illustrated in Figure~\ref{authentication_fig}. $U_i$ tries to prove herself to the verifier by executing the proposed authentication protocol shown in Figure~\ref{authentication_protocol}. Similar to the enrollment case, the biometric sensor must be authorized by the system before the authentication protocol is carried out. Steps of the proposed authentication protocol are as introduced as follows:

\begin{figure*} [t]
\centering
\fbox{\begin{tabular}{lcl}
{\sf \textbf{User ({\sffamily U}$_i$)}} & \hspace{-8cm} {\sf \textbf{Verifier ({\sffamily V})}}\\
\textbf{\small{Public:}} $pk_{U_i}$,$pk_i= (N_i,p_i^0,q_i^0)$  &\hspace{-8.5cm} \textbf{\small{Public:}} $pk_{U_i}$, $pk_i= (N_i,p_i^0,q_i^0$), \\ \hspace{8.5 cm}  $$\text{\sf Sign}$_{sk_{U_i}}(<C^{j}_{i}:j=1,\ldots,n>)$\\

\textbf{\small{Private:}} $sk_{U_i}, sk^1_i=(p_i^1, q_i^1), \textbf{\textit{B}}^j_{auth_i}={\textit{B}}^1_{auth_i} \ldots {\textit{B}}^n_{auth_i}$ & \hspace{-9.3cm} \textbf{\small{Private:}} $sk^2_i=(p_i^2, q_i^2)$\\\\

Choose $r^{j}_{i}\in_R \{0,1\}$ $\forall j=1, \ldots, n$   \\
Compute $R^{j}_{i}= r^{j}_{i}\oplus{\textit{B}}^j_{auth_i} \forall j=1, \ldots, n$\\
\\
\hspace{2cm}$\sends{<R^{j}_{i}:j =1,\ldots, n>, $nonce$_{U_i} }$ \hspace{1cm}\\
\\
& \hspace{-5.1cm} Retrieve $$\text{\sf Sign}$_{sk_{U_i}}(<C^{j}_{i}:j =1, \ldots, n>)$ \\
\\
\hspace{2cm}$\receives {$\text{\sf Sign}$_{sk_{U_i}}(<C^{j}_{i}:j =1, \ldots, n>), $nonce$_{V_i}}$\\
\\
Verify $$\text{\sf Sign}$_{sk_{U_i}}(<C^{j}_{i}:j =1, \ldots, n>) $\\
Compute ${C^{\prime}}^{j}_{i} = $\text{\sf Enc}$_{pk_i}(r^{j}_{i})\cdot C^{j}_{i}=\text{\sf Enc}_{pk_i}(r^{j}_{i}\oplus{\textit{B}}^j_{enroll_i}) \forall j=1, \ldots, n$\\
Compute $T^{1,j}_{i} = $\text{\sf Dec}$_{sk^1_i} ({C^{\prime}}^{j}_{i}) = ({C^{\prime}}^{j}_{i})^{(-p_i^1-q_i^1)/4} \mod N_i$ $\forall j =1, \ldots, n$\\
\\
\hspace{2cm}$\sends {$\text{\sf Sign}$_{sk_{U_i}}(<\text{\sf Enc}_{pk_i}(r^{j}_{i}),T^{1,j}_{i}:j =1, \ldots, n>, $nonce$_{U_i}, $nonce$_{V_i})}$\\
\\
& \hspace{-8cm} Verify $$\text{\sf Sign}$_{sk_{U_i}}(<\text{\sf Enc}_{pk_i}(r^{j}_{i}),T^{1,j}_{i}:j =1, \ldots, n>, $nonce$_{U_i}, $nonce$_{V_i})$\\
&\hspace{-13.6cm} Compute ${C^{\prime\prime}}^{j}_{i} = $\text{\sf Enc}$_{pk_i}(r^{j}_{i})\cdot C^{j}_{i}$\\
& \hspace{-9.7cm} Compute $T^{2,j}_{i} = $\text{\sf Dec}$_{sk^2_i} ({C^{\prime\prime}}^{j}_{i}) = ({{C^{\prime\prime}}^{j}_{i}})^{{(-p_i^2 - q_i^2)}/4} \mod N_i$ \\
& \hspace{-11.6cm} Compute $T^{3,j}_{i} = {{C^{\prime\prime}}^{j}_{i}}^{(N -p_i^0 -q_i^0 +1)/4} \mod N_i$ \\
& \hspace{-9.3cm} Compute decrypted bits $T^{j}_{i} = (1 - T^{1,j}_{i}T^{2,j}_{i}T^{3,j}_{i} \mod N_i)/2$ \\
& \hspace{-13.9cm} Compute $\sum^{k}_{j=1} R^{j}_{i} \oplus T^{j}_{i} \leq \mu$\\
\\
\hspace{4cm}$\receives {\text{Accept or Reject}} $\hspace {1cm}\\

\end{tabular}}

\caption{The Proposed Authentication Protocol}
\label{authentication_protocol}
\end{figure*}

\begin{enumerate}

\item \textbf{Step 1}: $U_i$ wants to verify her identity by using her biohash and sends a connection request to the verifier. Then, $U_i$ computes her biohash $\textbf{\textit{B}}_{auth_i}$ = ${\textit{B}}^1_{auth_i}$ $\ldots$ ${\textit{B}}^n_{auth_i}$ where ${\textit{B}}^j_{auth_i}$ $\in$ $\{0,1\}$, $j=1,\ldots,n$. Note that the user cannot produce exactly the same biometric template at each attempt and this results in different biohashes computed by the same user. Therefore, $\textbf{\textit{B}}_{enroll_i}$ and $\textbf{\textit{B}}_{auth_i}$ are different biohashes although they are generated by the same user at different sessions (enrollment and authentication). First, $U_i$ chooses a random vector $\textbf{\textit{r}}_i^j$ $\in_R$ $\{0,1\}$ $\forall$ $j=1,\ldots,n$. She computes $R^{j}_{i} = r^{j}_{i}\oplus{\textit{B}}^j_{auth_i}$ $\forall j=1,\ldots,n$. Then, $U_i$  generates a nonce,  nonce$_{U_i}$, which is uniquely defined and contains information about user id, session id and timestamp. Finally, the user sends $<R^{j}_{i}:j=1,\ldots,n>$, nonce$_{U_i}$ to the verifier.

\item \textbf{Step 2}: The verifier retrieves \text{\sf Sign}$_{sk_{U_i}}(<C^{j}_{i}:j=1,\ldots,n>)$ from the database where $C^{j}_{i}$ = $\text{\sf Enc}_{pk_i}\left(\textit{B}^j_{enroll_i}\right)$ $\forall$ $j=1,\ldots,n$. Then, it generates a nonce nonce$_{V_i}$ which contains information about the verifier, session id and timestamp. Finally, it sends \text{\sf Sign}$_{sk_{U_i}}(<C^{j}_{i}:j=1,\ldots,n>)$, nonce$_{V_i}$ to the user.

\item \textbf{Step 3}: The user verifies \text{\sf Sign}$_{sk_{U_i}}(<C^{j}_{i}:j=1,\ldots,n>)$ by using public key $pk_{U_i}$. She computes $C^{\prime j}_{i}$ = \text{\sf Enc}$_{pk_i}(r^{j}_{i})$ $\cdot$ $C^{j}_{i}$ = \text{\sf Enc}$_{pk_i}(r^{j}_{i}\oplus\textit{B}^j_{enroll_i})$ 
$\forall$ $j=1,\ldots,n$. Then, she performs partial decryption over $C^{\prime j}_{i}$, i.e., $T^{1,j}_{i}$ = \text{\sf Dec}$_{sk_i^1}(C^{\prime j}_{i})$ = ($C^{\prime j}_{i})^{(-p_i^1-q_i^1)/4}$ $\mod$ $N_i$, $\forall$ $j=1,\ldots,n$ using her private key share $sk_i^1$. Finally, she sends \text{\sf Sign}$_{sk_{U_i}}(<\text{\sf Enc}_{pk_i}(r^{j}_{i}), T^{1,j}_{i}:j=1,\ldots,n>$, nonce$_{U_i}$, nonce$_{V_i}$)  to the verifier.

\item \textbf{Step 4}: $V$ verifies the signature \text{\sf Sign}$_{sk_{U_i}}(<\text{\sf Enc}_{pk_i}(r^{j}_{i}), T^{1,j}_{i}:j=1,\ldots,n>$, nonce$_{U_i}$, nonce$_{V_i})$ by using the public key $pk_{U_i}$. Then, it computes ${C^{\prime\prime}}_i^j = $\text{\sf Enc}$_{pk_i}(r_i^j) \cdot C^{j}_{i}$ (this is done to assure correctness of the result and will prevent a malicious user computing different values than expected.). Next, the verifier performs the full decryption by computing $T^{2,j}_i= $\text{\sf Dec}$_{sk_i^2}$ $({C^{\prime\prime}}_i^j)$ = ${(C^{\prime\prime j}_{i})}^{(-p_i^2-q_i^2)/4}$ $\mod$ $N_i$ and $T^{3,j}_i$ = ${({C^{\prime\prime}}^{j}_{i})}^{(N-p_i^0-q_i^0+1)/4}$ $\mod$ $N_i$. Finally, the verifier computes the decrypted $j^{th}$ bits $T^{j}_{i}$ = $(1-T^{1,j}_{i}T^{2,j}_{i}T^{3,j}_{i} \mod N_i)/2$ and the Hamming distance  between $R^{j}_{i}$ and $T^{j}_{i}$ is calculated as follows:

\begin{equation}
\sum^{n}_{j=1}R^{j}_{i} \oplus T^{j}_{i} \leq \mu
\end{equation}
where $\mu$ is the distance threshold. Therefore, the verifier decides whether the user is authentic with respect to the pre-defined distance threshold. Note that the Hamming distance between $r^{j}_{i}\oplus\textbf{\textit{B}}^j_{enroll_i}$ and $r^{j}_{i}\oplus\textbf{\textit{B}}^j_{auth_i}$ is equal to the Hamming distance between $\textbf{\textit{B}}^j_{enroll_i}$ and $\textbf{\textit{B}}^j_{auth_i}$.

Finally, the verifier sends its decision (either Accept or Reject) to the user. However, the user may get dummy output if there is an error or an attack (i.e., override response attack) in the communication channel. The proposed system can easily be updated to cope with such an attack, for instance, by allowing the verifier to sign its decision including the nonces generated during the authentication session (i.e., either \text{\sf Sign}(Accept, nonce$_{U_i}$, nonce$_{V_i}$) or \text{\sf Sign} (Reject, nonce$_{U_i}$, nonce$_{V_i}$) and then sends it to the user. In this way, authenticity, integrity and origin of the data can easily be verified. Besides, signing the nonces (nonce$_{U_i}$ and nonce$_{V_i}$)  also makes the communication unique and avoids replay attacks.

\end{enumerate}


\textbf{Lemma 3.} \textit{Biohashes are not revealed at the authentication stage.}

\textbf{\textit{Proof.}} Authentication is performed in a randomized domain. In other words, the authentication is determined by comparing $R^j_i$ and $T^j_i$. An adversary can only obtain \textit{R}$^j_i$ and \textit{T}$^j_i$ which are revealed at the authentication stage. Recall that these are randomized biohashes. Thus, from the adversary's perspective, there are three unknowns $(\textit{r}_i^j$, $\textit{B}^j_{enroll_i}$ and $\textit{B}^j_{auth_i})$ and two equations which are shown in the below. 

\begin{equation}  
\textit{T}_i^j = \textit{r}_i^j\oplus\textit{B}^j_{enroll_i}
\end{equation}

\begin{equation}
\textit{R}_i^j = \textit{r}_i^j\oplus\textit{B}^j_{auth_i}
\end{equation}
where $r_i^j$ is the random bit generated by the $U_i$ for the $j^{th}$ bit. Since this is a system of linear equations with fewer equations than unknowns, the system has infinitely many solutions. Consequently, it is impossible for the adversary to obtain a legitimate user's biohash by using $T_i^j$ and $R_i^j$ which are revealed at the authentication stage. As a result, the proposed biometric authentication system ensures security and privacy. 

\section{Security proof of the proposed authentication protocol}

In this section, we prove that the proposed authentication protocol shown in Figure~\ref{authentication_protocol} is secure against malicious user and verifier. In this proof, there exists a probabilistic polynomial-time simulator that produces a protocol transcript which is statistically indistinguishable from the one resulting from a real execution of the authentication protocol. The simulator must perform its task without knowing the private information of the party who proves her identity \cite{Goldreich}. We show that given a party is corrupted (either a user or verifier), there exists a simulator that can produce a view which is statistically indistinguishable from the view of that party interacting with the other honest party. Assuming that one party is corrupted, we build an efficient simulator that has access to the public input and  private secret shares of the secret key  of the corrupted party. Besides, the simulator knows the public output. We want to point out that the simulator already knows the shares of the secret key of the corrupted party before the simulation is run. Since the threshold cryptosystem is set up before the protocol starts, we assume that the simulator extracts this information when the distributed key generation is run.

It is worth mentioning that the proposed authentication protocol gives computational privacy to both the user and the verifier due to the semantic security of the underlying cryptosystem. Furthermore, it is shown that the proposed authentication protocol is simulatable for both parties and these simulations produce views which are statistically indistinguishable from the views in the real protocol executions.

\textbf{Theorem 1.} \textit{The proposed authentication protocol, which is shown in Figure~\ref{authentication_protocol}, is secure in the presence of static malicious adversaries.}

\textbf{\textit{Proof.}} We show that given a party is corrupted, there exists a simulator that can produce a view to the adversary that is statistically indistinguishable from the view in the real protocol execution based on its private decryption share as well as public information.

{\it Case 1 - User $U_i$ is corrupted}. In this case, we prove the security for the case where $U_i$ is corrupted. The simulator has the private key share of the user $sk_i^1$, the user's private key $sk_{U_i}$, and the user's biohash $B^j_{auth_i}$ apart from the user's public information (i.e., $pk_{U_i}$ and $pk_i$) as described in the proposed authentication protocol. The simulator constructs a view for the user which is statistically close to the one the user observes when interacting with the honest verifier by using this information. The simulator proceeds as follows:

\begin{enumerate}

\item The simulator first obtains $<R^{j}_{i}:j =1,\ldots, n>, $nonce$_{U_i}$. As in the second round of the real protocol, the simulator needs to output the signature of the encrypted biohash of the user. To do so, the simulator computes $\tilde{C}^{j}_{i} = \text{\sf Enc}_{pk_i}(B^j_{auth_i})$ $\forall$ $j=1,\ldots,n$ by using the user's public key $pk_i$, and then computes \text{\sf Sign}$_{sk_{U_i}}\left(\left\langle \tilde{C}^{j}_{i}:j=1,\ldots,n\right)\right\rangle$. The simulator also generates a nonce called $\tilde{\text{nonce}}$$_{V_i}$.  The values \text{\sf Sign}$_{sk_{U_i}}\left(\left\langle \tilde{C}^{j}_{i}:j=1,\ldots,n\right)\right\rangle$ and $\tilde{\text{nonce}}$$_{V_i}$ are the simulated outputs.  Note that the simulator uses $\tilde{\textbf{B}}_{auth}$ instead of \textbf{B}$_{enroll}$ since it is the only available biohash to him.

\item The simulator obtains \text{\sf Sign}$_{sk_{U_i}}(<\text{\sf Enc}_{pk_i}(r_i^j),T^{1,j}_i:j =1, \ldots, n>, $nonce$_{U_i}, \tilde{\text{nonce}}_{V_i})$ as in the second round of the protocol.  The simulator next verifies the signature \text{\sf Sign}$_{sk_{U_i}}(<\text{\sf Enc}_{pk_i}(r_i^j),T^{1,j}_i:j =1, \ldots, n>, $nonce$_{U_i}, \tilde{\text{nonce}}_{V_i})$ that $U_i$ would run. Next, it computes $\tilde{C}^{\prime \prime j}_i = $\text{\sf Enc}$_{pk_i}(r_i^j)\cdot \tilde{C}_i^j$. 

Given $\tilde{C}^{\prime \prime j}_{i}$, its plaintext and the share of private key $sk_i^2$ of the user $U_i$ the decryption shares $T_i^{2,j}$ can be simulated as follows: The simulator computes $\tilde{b}_{0} = [\tilde{C}^{\prime \prime j}_{i}]^{{(N - p_{0} - q_{0} +1)}/4} \mod N$ from the public information and computes $\tilde{b}_{1} = [\tilde{C}^{\prime \prime j}_{i}]^{{(-p_{1}-q_{1})}/4} \mod N$ since it knows $sk_i^1$ (i.e., $p_1, q_1$). Let's denote $\tilde{b}$ for the plaintext of $\tilde{C}^{\prime \prime j}_{i}$. Then, the simulator can compute $\tilde{b}_2 \mod N \equiv (1- \tilde{2b})/(\tilde{b}_0 \tilde{b_1}) \mod N$ (which is $T_i^{2,j}$ in the real protocol). Note that in the real setting this is not possible since the plaintext inside the ciphertext is unknown $\tilde{C}^{\prime \prime j}_{i}$.

Similarly, $T_i^{3,j}$ can also be simulated since $p_i^0$ and $q_i^0$ are known by the simulator. The simulator finally computes $\sum^{k}_{j=1} R_i^j \oplus T_i^j$. 

\end{enumerate}

Each step of the proposed authentication protocol for the simulator is simulated and this completes the simulation for the malicious user. The transcript is consistent and statistically indistinguishable from the user's view when interacting with the honest verifier. 

{\it Case 2 - The verifier $V$ is corrupted}. We now prove the security for the case where the verifier is corrupted. The simulator has the private key share of the verifier $(sk_i^2)$ apart from the verifier's public information ( i.e., $pk_{U_i}$, $pk_i$, and \text{\sf Sign}$_{sk_{U_i}}(<C_i^j:j=1,\ldots,n>))$ as described in the proposed authentication protocol. The simulator constructs a view for the verifier which is statistically close to the one when interacting with the honest user by using this information. The simulator proceeds as follows:

\begin{enumerate}

\item Note that the simulator already knows $<C_i^j:j=1,\ldots,n>$ because of the knowledge of \text{\sf Sign}$_{sk_{U_i}}(<C_i^j:j=1,\ldots,n>)$. The simulator chooses a random bit $\tilde{r}_i^j$ and arbitrary $\tilde{B}^j_{auth_i}  \in_R \{0,1\}$ and computes $\tilde{R}_i^j = \tilde{r}_i^j \oplus \tilde{B}^j_{auth_i}$. Recall that the simulator must perform its task without knowing the private information of the honest user in this case. Thus, although it does not have real $r_i^j$ and $B^j_{auth_i}$ it can successfully execute the simulated conversation since $\tilde{R}_i^j$ is uniformly random.

\item The simulator obtains \text{\sf Sign}$_{sk_{U_i}}(<C_i^j:j=1,\ldots,n>)$. The simulator verifies the signature \text{\sf Sign}$_{sk_{U_i}}(<C_i^j:j=1,\ldots,n>)$ (by using the user's public key $pk_{U_i}$) that $V$ would run. Next, it next computes $\tilde{C}^{\prime j}_{i} = \text{\sf Enc}_{pk_i}(\tilde{r}_i^j) \cdot C_i^j$. 

\item Given $\tilde{C}^{\prime j}_{i}$, its plaintext (which is $\tilde{r}_i^j \oplus B^j_{enroll_i}$) and the share of private key $sk_i^2$ of the verifier $V$ the decryption share $\tilde{T}_i^{1,j}$ can also be simulated as follows: The simulator computes $\tilde{b}_0= [\tilde{C}^{\prime j}_{i}]^{{(N -p_0 -q_0 +1)}/4} \mod N$ from the public information and computes $\tilde{b}_2 = [\tilde{C}^{\prime j}_{i}]^{{(-p_2-q_2)}/4} \mod N$ since it knows $sk_i^2$ (i.e., $p_2, q_2$). Let's denote $\tilde{b}$ for the plaintext of $\tilde{C}^{\prime j}_{i}$. Then, the simulator can compute $\tilde{b}_1 \mod N \equiv (1- \tilde{2b})/(\tilde{b}_0 \tilde{b_2}) \mod N$ (which is $T_i^{1,j}$ in the real protocol). Note that in the real setting this is not possible since the plaintext inside the ciphertext is unknown $\tilde{C}^{\prime \prime j}_{i}$.

\item Finally, the simulator needs to simulate the signature $\text{\sf Sign}_{sk_{U_i}}(<\text{\sf Enc}_{pk_i}(r^{j}_{i}),T^{1,j}_{i}:j =1, \ldots, n>, $nonce$_{U_i}, $nonce$_{V_i})$. However, this is not possible since the simulator does not know $sk_{U_i}$. In order to successfully simulate this final step, we need to provide additional information to the simulator. For example, performing a following modification over the the key generation phase in the real protocol the simulation will be possible: 

\begin{itemize}
\item During the key generation in the real protocol, the private key $sk_{U_i}$ is distributed to the user $U_i$ and $V$ in a threshold fashion. For example, distributed RSA setting can be used for the signature algorithm (note that for an RSA setting $(e,n)$ denotes the public key and $(p,q,d)$ denotes the private key where $n = pq$ and $ed \equiv 1 \mod (p-1)(q-1)$). Namely, the private key $d$ can be divided into $d_1$ and $d_2$ such that the ciphertext $c$ can be decrypted together with $U_i$ and $V$ as $m$ $\equiv$ $c^d$ $\equiv$ $c^{d_1+d_2} \mod n$. 

\item In order to simulate, instead of signing procedures, $U_i$ will compute an encryption, compute its partial decryption and finally will send to the verifier. This will assure that the user indeed used its private decryption key over the encrypted value. Next, the verifier will also compute its partial decryption and will compute the decrypted value privately.
\end{itemize}

Hence, with this modified version the decryption share can be simulated in a similar way as described at the third step of the simulation. 

\end{enumerate}

Consequently, each step of the proposed authentication protocol for the simulator is simulated and this completes the simulation for the malicious verifier. The transcript is consistent and statistically indistinguishable from the verifier's view when interacting with the honest user. 

\qed

\section{Complexity Analysis of the Proposed System}

In this section, we discuss the complexity of the THRIVE enrollment and authentication protocols. The complexity of the THRIVE enrollment and authentication protocols are examined in terms of protocol steps for the round complexity, the number of cryptographic operations for the computational complexity and the number of messages exchanged by the two parties for the communication complexity. Without loss of generality, we will provide complexity of the THRIVE protocols using the (2,2)-threshold homomorphic GM cryptosystem as an instance~\cite{GM84}. In the protocol we use (2,2)-threshold XOR homomorphic GM cryptosystem for confidentiality (i.e., encryption and decryption) while for signature generation and verification a conventional cryptosystem such as RSA (using the key pair $(pk_{U_i},sk_{U_i})$) is employed.

The round complexity of the enrollment protocol is only one. For the computational complexity, the enrollment protocol requires $n$ XOR-homomorphic encryptions, and one conventional signature generation for a user, but one signature verification for the server. For the communication complexity, the user sends a conventional signature and $n$ ciphertexts (i.e., $C_i^j$ for $j = 1, \ldots n$). 

In the authentication protocol, there are only four rounds. For the computational complexity of the authentication protocol, the user generates one conventional signature and verifies another, computes $n$ XOR-homomorphic encryptions and $n$ XOR-homomorphic decryptions, and performs $n$ modular multiplications over homomorphic ciphertexts (i.e., $Enc_{pk_i}(r_i^j)\cdot C_i^j$ for $j = 1, \ldots n$). The verifier verifies one conventional signature, computes $n+2$ modular multiplications, $2n$ decryptions, and performs $n$ Jacobi computations to check $Enc_{pk_i}(r_i^j)$ for  $j = 1, \ldots n$. In total, there are $n$ XOR-homomorphic encryptions, $3n$ XOR-homomorphic decryptions, two signature verifications, one signature generation, $n$ Jacobi computations, and $2n+2$  modular multiplications during the entire authentication protocol.

\begin{figure}[tb]
\begin{center}
\includegraphics [scale=0.47]{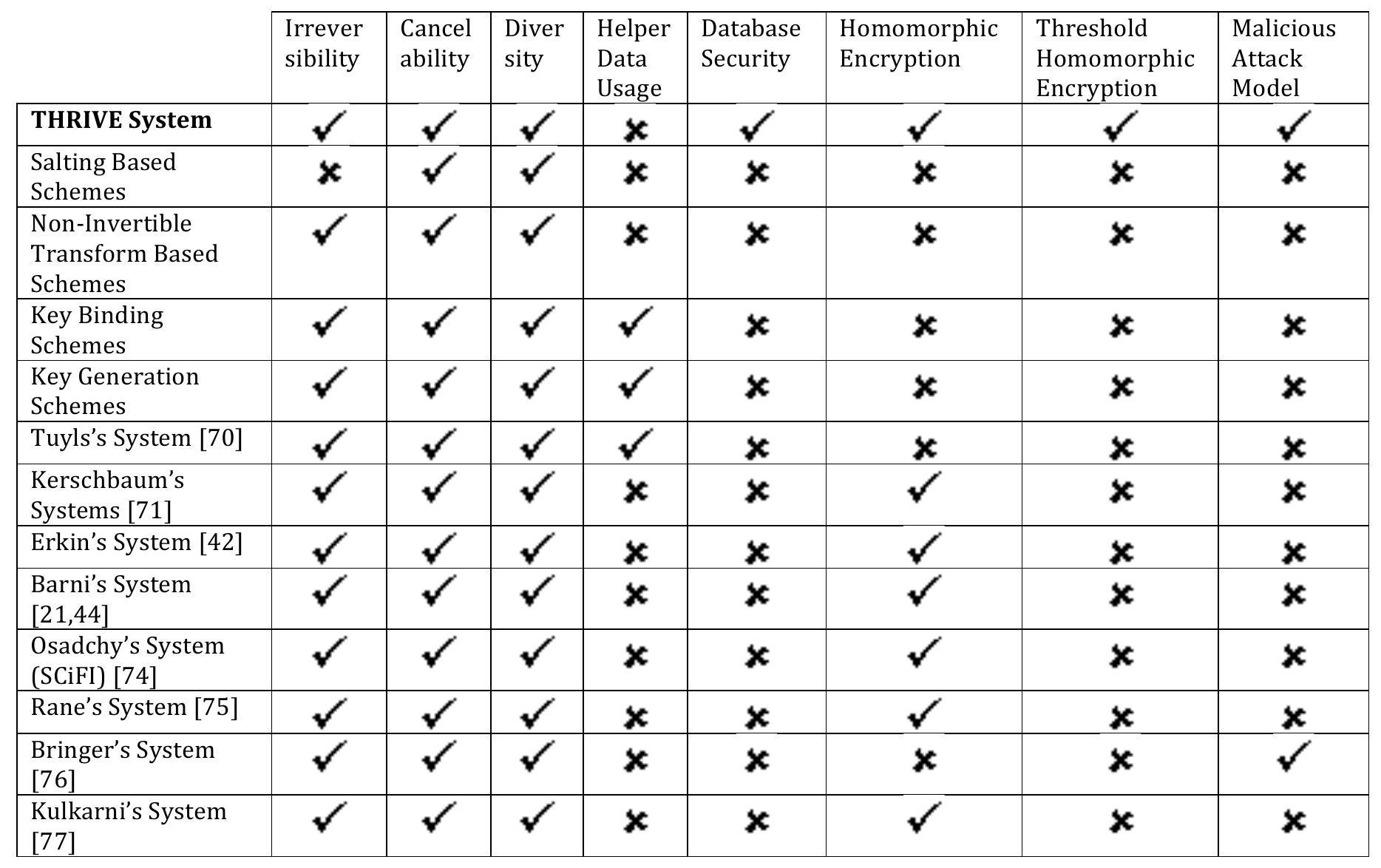}
\end{center}
\caption{Comparison between the THRIVE system and the existing solutions. \ding{51} denotes that the system satisfies the property whereas \ding{55} denotes that the system does not satisfies the property.}
\label{compare}
\end{figure}  

For the communication complexity of the authentication protocol, the user sends $2n$ homomorphic ciphertexts, one conventional signature and one nonce value. The verifier sends one conventional signature, $n$ homomorphic ciphertexts, and one nonce to the user. In total, $3n$ homomorphic ciphertexts, two conventional signatures, and two nonce values are exchanged. 

In the following, we provide timing estimates for the entire protocol for 80-bit security level, on a desktop computer, which has Intel processor with various clock speeds (2.4 GHz and 3.2 GHz). On the computer one modular multiplication using Montgomery arithmetic takes about 2000 clock cycles. Furthermore, one decryption operation in XOR homomorphic GM cryptosystem takes only a single modular multiplication. On the other hand, one decryption in XOR homomorphic GM cryptosystem requires one modular exponentiation operation which takes about $3.2$ million clock cycles. In addition, one signature generation and verification operation in conventional cryptosystem such as RSA are equivalent to one modular exponentiation operation. 

The bandwidth usage of the proposed protocol for various lengths of biohashes of the user are given in Table ~\ref{bandwidth}. The required bandwidth for the proposed protocol increases with the increasing length of the biohashes. Bandwidth usage also affects the overall connection time.

\begin{table}[h]
\caption{Bandwidth (total number of bits exchanged) usage of the proposed protocol.}
\label{bandwidth}
   \begin{tabular}{|p{2.3cm}|p{2.3cm}|p{2.8cm}|} \hline
    Length of Biohash  & Bandwidth (Kbits) &                Time @ 10 Mbit/s (ms) \\ \hline
    112                & 348               & 35                                   \\ \hline
    192                & 594               & 59                                   \\ \hline
    256                & 791               & 79                                   \\ \hline
    512                & 1577              & 158                                  \\ \hline
    2048              & 6296              & 630                                  \\ \hline
    \end{tabular}
\end{table}

The computation times for the user and the verifier at 2.4 GHz for the proposed protocol with different biohash lengths are given in Table ~\ref{user_time1}. Naturally, it is expected that the required computation times of the proposed protocol increase as lengths of the biohashes increase.

\begin{table}[h]
    \caption{Computation time for the user and the verifier at 2.4 GHz.}
\label{user_time1}
\begin{tabular}{|p{2.3cm}|p{2.5cm}|p{2.5cm}|}
    \hline
    Length of Biohash & User Time (ms) & Verifier Time (ms) \\ \hline
    112               & 151                          & 449                       \\ \hline
    192               & 258                          & 769                       \\ \hline
    256               & 343                          & 1026                       \\ \hline
    512               & 685                          & 2050                     \\ \hline
    \end{tabular}
\end{table}

We compare the communication complexity of the proposed system with the existing systems in the literature assuming that all systems run on a computer platform with 2.4~GHz clock speed. Erkin \textit {et al.}'s system \cite{Erkin} requires 56.25 ms, Barni \textit {et al.}'s system \cite{Barni, Barni2} requires 50 ms and Sadeghi \textit {et al.}'s system \cite{Sadeghi} requires 25 ms for authentication at the server side for single user with 112~bit binary feature vector \cite{Barni}. On the other hand, our solution requires 449 ms for the same authentication set-up. Although existing solutions seem faster than the proposed system, they propose their solutions in the semi-honest model whereas our solution is secure under malicious adversary model. The comprehensive comparison between the proposed system and existing systems are shown in Figure ~\ref{compare}. It is clearly seen in Figure ~\ref{compare} that the proposed THRIVE system offers superior security and privacy preservation solutions under malicious attack model.

The similar timing estimations are also computed for 3.2~GHz as shown in Table ~\ref{user_time2}.

\begin{table}[h]
    \caption{Computation time for the user and the verifier at 3.2 GHz.}
\label{user_time2}
\begin{tabular}{|p{2.3cm}|p{2.5cm}|p{2.5cm}|}
    \hline
    Length of Biohash & User Time (ms) & Verifier Time (ms) \\ \hline
    112               & 113                          & 337                      \\ \hline
    192               & 193                          & 577                      \\ \hline
    256               & 257                          & 769                       \\ \hline
    512               & 514                          & 1537                      \\ \hline
    2048             & 2051                         & 6146                      \\ \hline
    \end{tabular}
\end{table}

We compare the communication complexity of the proposed system with the existing systems at 3.2 GHz in the literature. Kulkarni \textit {et al.}'s system \cite{Kulkarni} requires 58~s at the server side, 10 ms at the user side and 400 Kbit bandwidth usage for authentication of  single user with 2048-bit binary feature vector at 3.2 GHz. Our proposed system requires 6146 ms at the server side, 2051 ms at the user side and 6296 Kbit bandwidth usage for the same authentication set-up. Thus, it is faster than Kulkarni \textit {et al.}'s system. In addition, our proposes system offers other advantages such as that it is secure under malicious adversary model, that the biometric is protected via both a template protection method (e.g., biohash) and cryptographic primitives (e.g. threshold homomorphic encryption). On the other hand, Kulkarni \textit {et al.}'s system offers solution for semi-honest model which also means that it is insecure for the malicious model. The comprehensive comparison between the proposed system and existing systems are shown in Figure ~\ref{compare}. 

\section{Conclusion}

In this work, we propose a novel biometric authentication system. The aim of the proposed system is to increase security against adversary attacks defined in \cite{Ratha2} when an adversary wants to gain access to the system as a legitimate user. We propose novel enrollment and authentication protocols to increase the security against attacks reported in the literature and preserve the privacy of users. The proposed system can be used with any biometric feature extraction method which can produce binary templates or whose outputs can be binarized. The biohashing is chosen as an example binary biometric template generation system since it offers satisfactory error rates and fast authentication. The comparison is performed in a randomized domain in the authentication stage and the binary templates (e.g., biohashes) are never released. In addition, only encrypted binary templates are stored in the database. Since we use the $(2,2)$-threshold cryptosystem, the verifier cannot decrypt the data stored in the database by himself. Namely, the user and the verifier both has to cooperate to decrypt the encrypted binary templates. The proposed system can be used in applications where the user is not willing to reveal her biometrics to the verifier although she needs to proof her physical presence by using biometrics. The THRIVE system is suitable for applications where the user and the verifier do not necessarily trust each other. Furthermore, the proposed system appears to be sufficiently efficient compared to the existing scheme and can be used in real-life applications.

\section{Acknowledgments}

This work has been performed by the BEAT project $7^{th}$ Framework Research Programme of the European Union (EU), grant agreement number: 284989. The authors would like to thank the EU for the financial support and the partners within the consortium for a fruitful collaboration. For more information about the BEAT consortium please visit http://www.beat-eu.org.

\bibliographystyle{IEEEtran}
\bibliography{references}

\end{document}